\documentclass[english,manuscript]{revtex4}
\usepackage{times}
\usepackage[T1]{fontenc}
\usepackage[latin1]{inputenc}
\usepackage{graphicx}

\makeatletter

\usepackage{babel}
\makeatother
\begin{document}

\title{From Graphene constrictions to single carbon chains}

\author{Andrey Chuvilin}

\thanks{These authors contributed equally to this work}

\author{Jannik C. Meyer}

\thanks{These authors contributed equally to this work}

\author{Gerardo Algara-Siller}

\author{Ute Kaiser}

\affiliation{Electron microscopy group of materials science, University of Ulm,
Germany}

\begin{abstract}
We present an atomic-resolution observation and analysis of graphene
constrictions and ribbons with sub-nanometer width. Graphene membranes
are studied by imaging side spherical aberration-corrected transmission
electron microscopy at 80 kV. Holes are formed in the honeycomb-like
structure due to radiation damage. As the holes grow and two holes
approach each other, the hexagonal structure that lies between them
narrows down. Transitions and deviations from the hexagonal structure
in this graphene ribbon occur as its width shrinks below one nanometer.
Some reconstructions, involving more pentagons and heptagons than
hexagons, turn out to be surprisingly stable. Finally, single carbon
atom chain bridges between graphene contacts are observed. The dynamics
are observed in real time at atomic resolution with enough sensitivity
to detect every carbon atom that remains stable for a sufficient amount
of time. The carbon chains appear reproducibly and in various configurations
from graphene bridges, between adsorbates, or at open edges and seem
to represent one of the most stable configurations that a few-atomic
carbon system accomodates in the presence of continuous energy input
from the electron beam.
\end{abstract}
\maketitle
Carbon is one of the most important elements that occurs in numerous
allotropes, displays an exceedingly rich chemistry, and is contained
in a staggering number of compounds. The two solid crystalline forms,
graphite and diamond, are known since ancient times, while the more
recently discovered fullerenes \cite{C60NobelPaper85}, carbon nanotubes
\cite{IijimaMWNT1991,Dresselhaus_PhysicalPropsOfCNTs1998}, and graphene
\cite{Novoselov2004Sci,GeimGrapheneReview07} make up a large part
of today's nanotechnology research. Thus, a wide range of allotropes
and structural species with all dimensionalities is available for
research and applications today. Now, we present a simple and reliable
synthesis of a truly one-dimensional carbon species, a single-atomic
linear carbon chain. Evidence of such carbon chains has been observed
in spectral signatures from space \cite{CChainSmalleySpace87} and
in laboratory experiments \cite{CChainSmalley87,CChainAtCntNL03,CChainCumuleneSci08};
however their formation mechanisms have remained unclear and the chemical
and physical properties mostly undiscovered. Here, we form monoatomic
carbon chains by shrinking a graphene constriction under electron
irradiation. We follow their formation with atomic resolution and
sufficient sensitivity to detect every carbon atom that remains stable
for at least one second. We find that graphene constrictions deviate
from the hexagonal structure as their width is reduced below one nanometer,
and intermediate structures between a graphene constriction and a
carbon chain are dominated by pentagons and heptagons. The carbon
chain formation from shrinking a graphene constriction indicates a
possible site specific fabrication, which is the first step to a further
analysis or technological application.

Our study of graphene membranes reveals an efficient formation mechanism
of carbon chains. Indeed, these chains appear almost-inevitably in
these ultra-thin graphitic samples after a sufficiently high dose
of electron irradiation. Holes and constrictions form in the process
of irradiation and, most surprisingly, most of the constrictions turn
into carbon chains before ultimately detaching. Thus, instead of a
synthesis from smaller units, the carbon chains form by self-organization
from a continuously diluted set of carbon atoms in the presence of
ionizing irradiation. In addition, the amorphous, carbonaceous adsorbates
on the graphene membranes form carbon chains while shrinking under
electron irradiation, and bended chains appear at the free edges of
graphene. This variety of conditions in which they form indicates
that these chains are a preferential and stable configuration at a
low density of carbon atoms, not limited to graphene membranes. Further,
by shrinking a graphene ribbon \cite{GraNREnGapsLouie07} in an electron
beam under continuous observation, we investigate the transition from
a quasi-2D material to a 1D structure. As the width drops below ca.
1 nanometer, the structural behaviour of the ribbon and its two edges
becomes distinctively different from that of a semi-infinite sheet
with one open edge \cite{GraAtEdge08,GraEdgeReconstPRL08,GraEdgeStressPRB08},
and thus marks the transition from a surface-dominated to an edge-dominated
regime. Stable planar $sp^{2}$-bonded reconstructions of the graphene
bridge are observed as intermediate configurations. 

Graphene membranes are prepared by mechanical cleavage of graphite
and transfer to commercially available transmission electron microscopy
(TEM) grids as described previously \cite{MeyerEBIDapl08}. The presence
of a single layer is verified by electron diffraction \cite{MeyerGrapheneSSC07}.
TEM investigations are carried out using an imaging-side spherical
aberration corrected Titan 80-300 (FEI, Netherlands), operated at
80kV. The electron beam current density is ca. $3\cdot10^{7}\frac{e^{-}}{\textrm{s}\cdot\textrm{nm}^{2}}$.
The spherical aberration is set to 20\ensuremath{µ}m and imaging is
done at Scherzer conditions \cite{Spence_HRTEM}. The extraction voltage
of the field emission gun is reduced from its standard value (3.8kV)
to 1.7kV in order to reduce the energy spread of the source. This
results in a clear improvement of contrast and resolution, both of
which are limited by chromatic aberrations of the objective lens at
80kV operating voltage. 

A key advantage of spherical aberration corrected TEM is that the
point resolution can be set approximately to the information limit
of the microscope. In this way, effects of delocalization are reduced
\cite{UrbanCsCorrOptHRTEM02}. Thus, for a sample that is imaged at
optimum focus conditions and that is thin enough to neglect multiple
scattering, the images can be directly interpreted in terms of the
atomic structure. For a single layer of carbon, these requirements
are easily fulfilled. Atoms appear black at our conditions.

Holes appear in graphene membranes during electron irradiation, as
described previously \cite{GraHoleCutAPL08,GraAtEdge08}. We look
for configurations where two nearby holes are present in a clean region
of the graphene sheet. Although we look here for the coincidental
formation of two nearby holes, we note that it has been demonstrated
by Fischbein et al. that an arbitrary configuration of holes can be
made in a controlled position by electron irradiation \cite{GraHoleCutAPL08}.
At an electron energy of 80kV, atoms at the graphene edges are removed
while the continous graphene membrane areas are very stable. Thus,
as the holes grow under continuous electron irradiation, the graphene
bridges between nearby holes inevitably shrink but are not damaged
otherwise. Eventually, narrow graphene constrictions form between
adjacent holes. We record a continuous sequence of images on the CCD
camera, using an exposure time of 1 second at 4 second intervals.
Reconstructions that affect not only the edges but the entire graphene
ribbon are frequently observed at widths of less than 1 nanometer.
Finally, a single chain with a typical length of 10-15 carbon atoms
is seen in more than 50\% of the cases as the final product of bridge
thinning, which then remains stable for up to two minutes under our
intense electron beam. 

\begin{figure}
\includegraphics[width=0.64\linewidth]{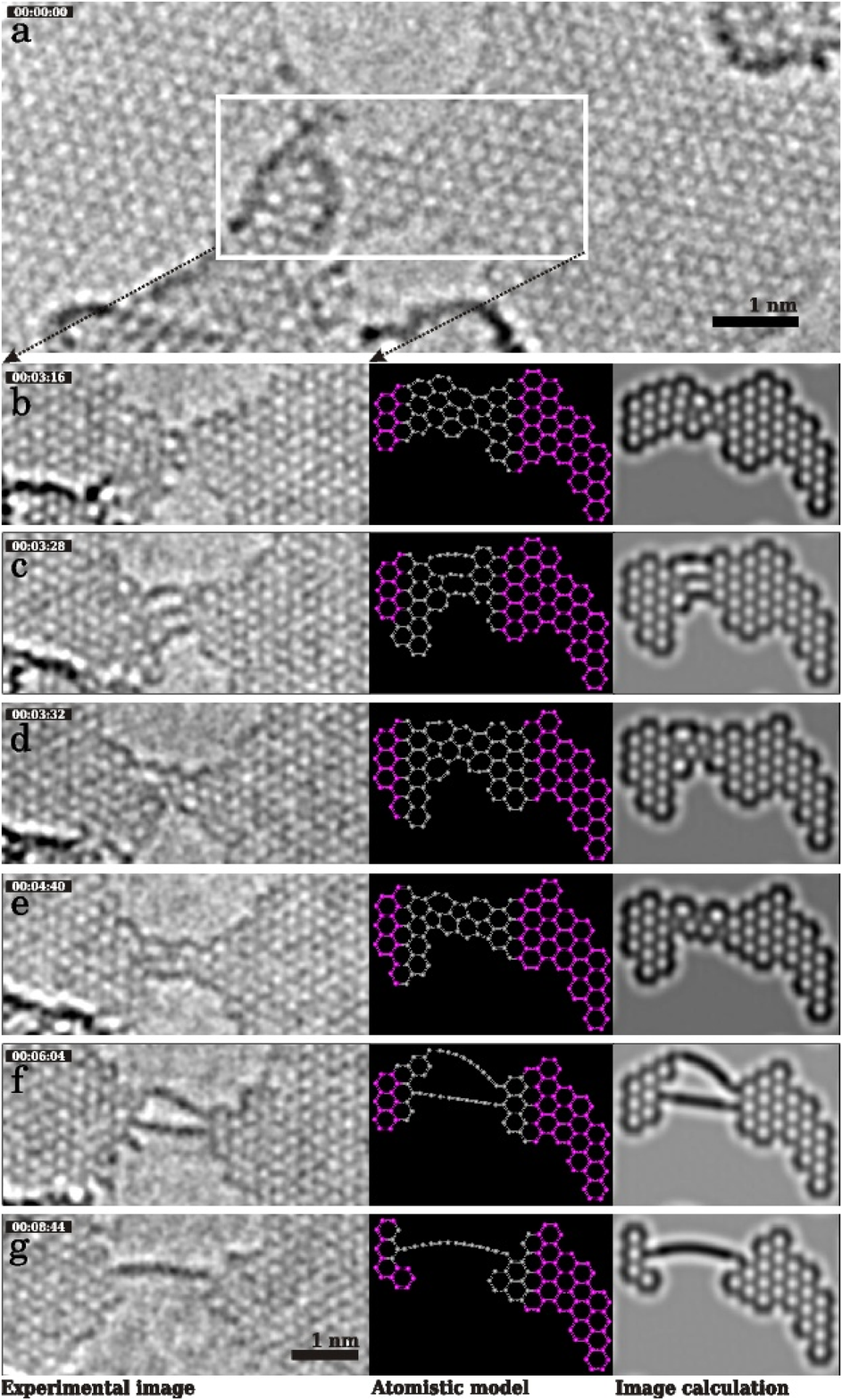}

\caption{Transition from a graphene ribbon to a single carbon chain. (a) HRTEM
image of the initial graphene ribbon configuration (atoms appear black).
In this image, the graphene ribbon runs horizontally. A tiny adsorbate
is present on the left hand side of the bridge, but it disappears
shortly afterwards. (b-g) Time evolution of the bridge, in the experimental
image (left), atomistic model (center), and corresponding image calculation
(right). Carbon chains are present in panels (c, f, g); reconstructed
bridge configurations of planar covalent carbon networks are seen
in (b, d, e). Note that the higher contrast of the carbon chains compared
to the graphene lattice is in agreement with the image calculation.
\label{fig:bigimage}}
\end{figure}

A single layer graphene membrane with two holes, separated by a ca.
1nm wide graphene bridge, is shown in Fig. \ref{fig:bigimage}a. Figs.
\ref{fig:bigimage}b-g show the same graphene constriction at later
times, along with best-fit atomic models and image simulations. A
video of this process is shown in the supplementary video S1 (supplementary
videos S2, S3 show the time evolution of two similar graphene constrictions).
Again, the atomic model is easily derived because the dark contrast
in this spherical aberration corrected image can be directly interpreted
in terms of the atomic structure. Further, the structure appears to
remain planar, i.e. no indication of two carbon atoms on top of each
other in the projected structure was observed (small out-of-plane
distortions are still possible, and indeed likely, in a carbon structure
with multiple pentagons and heptagons). The atomistic model is generated
with the ghemical software \cite{Ghemical}, and the TEM image calculation
is obtained using multislice software \cite{MusliSWultramicr05}.

The reconstructions that occur in the time evolution, with continuous
energy input from the electron beam, are indeed remarkable. In this
example, the graphene bridge incorporates pentagons and heptagons
(Fig. \ref{fig:bigimage}b), converts into three parallel carbon chains
(Fig. \ref{fig:bigimage}c), transforms back into a structure with
multiple polygons (5, 6, 7, 8 sides, Fig. \ref{fig:bigimage}d, e),
and finally becomes a double and single carbon chain before the two
holes merge into one (Fig. \ref{fig:bigimage}f, g). Further, it can
be seen in the supplementary video S1 that the end point of the carbon
chain, i.e. its connection to the hexagonal mesh of the graphene sheet,
drifts between nearby edge atoms multiple times before the chain finally
breaks. Similar transitions or reconstructions were observed many
times in different samples, and are separately discussed in the following
paragraphs.

\begin{figure}
\includegraphics[width=0.5\linewidth]{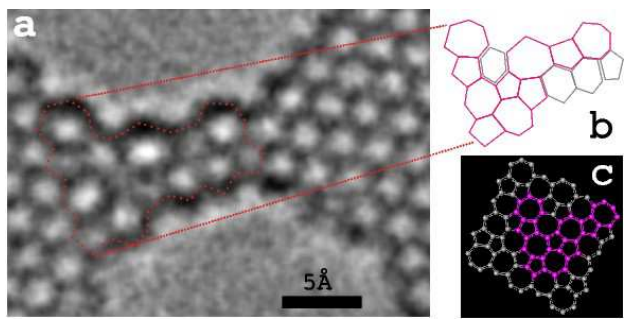}

\caption{A stable configuration discovered within the reconstructions of a
graphene bridge. (a) HR-TEM image (atoms appear black). (b) Atomistic
model of the outlined region. (c) A model of pentaheptite \cite{PentaheptidePRB96}.
Matching atom configurations in the experiment (b) and pentaheptite
model (c) are marked in pink.\label{fig:pentaheptide}}
\end{figure}

We begin our discussion with the transformations from a hexagonal
structure to meshs that are dominated by pentagons and heptagons.
From the observed structures alone, it appears that a variety of (seemingly
random) combinations of pentagons, heptagons or higher polygons can
be generated within the graphene bridge. The reason for these reconstructions
may be twofold: As an intrinsic origin, calculations of edge stress
\cite{GraEdgeStressPRB08} and edge reconstructions \cite{GraEdgeReconstPRL08}
indicate that unperturbed graphene edges might not be the optimum
configuration. Thus, reconstruction of the entire structure can be
expected once the graphene ribbon is sufficiently narrow. As an extrinsic
driving force, energy input from the electron beam is present, which
can help e.g. to overcome activation barriers. Indeed, the dominant
effect of the electron irradiation at 80kV in graphene is not atom
removal, but atom rearrangement (e.g. the Stone-wales type bond rotation
\cite{StoneAndWalesArticleCPL86}). Thus, structures that are random
(to some extent) are formed locally. At the same time, these structures
are observed with single-atom precision. We can now gain crucial insights
by studying the stability (under the beam) of the observed structures,
since {}``stable'' configurations would be expected to last longer
than unstable ones. For example, we find that the configuration of
Fig. \ref{fig:bigimage}e turns out to be stable for 2 minutes in
the intense electron beam, which corresponds to an electron dose of
$\approx4\cdot10^{9}\frac{e^{-}}{\textrm{nm}^{2}}$ and is longer
than all other multiple-polygon type structures observed within our
data. Fig. \ref{fig:pentaheptide}a shows an image with better signal-to-noise
ratio (by averaging of 10 CCD frames) of this structure. From the
atomic configurations (Figs \ref{fig:pentaheptide}b,c), it turns
out that the structure is locally that of pentaheptite, a carbon allotrope
that was predicted to be stable in 1996 \cite{PentaheptidePRB96}
but never observed experimentally so far. 

As a generalization, it appears promising to look for stable configurations
with the continuous {}``randomization'' of some atoms by the electron
beam. In this way, configurations with local energetic minima can
be discovered (such as pentaheptite) which are not experimentally
accessible otherwise. Graphene bridges provide an ideal starting point
for such a study: Bond rotations were observed previously in graphene
without holes but then they are constrained by the continuous membrane,
and thus relax to the unperturbed structure \cite{MeyerTEAMnl08}.
Graphene edges, the boundary of a semi-infinite sheet, allow to study
the migration of edge atoms but still the reconstructions do not seem
to penetrate into the graphene sheet \cite{GraAtEdge08} (topological
changes are limited to the edge polygons). Graphene bridges or ribbons
are only constrained in one direction and can reconstruct more freely,
but at the same time the structures remain sufficiently planar and
stable for a direct analysis by low voltage, aberration corrected
transmission electron microscopy. Thus, we expect that a further study
of intermediate structures in graphene bridges, with particular respect
to their stability, can provide further insights on the complicated
bonding behaviour in carbon materials. 

We now turn to the chain-type configurations that are frequently observed
as the graphene bridge is ultimately narrowed down to a few-atom or
single-atom width. It is obvious that with only a random removal of
atoms, even of only the edge atoms from a graphene bridge, the formation
of a carbon chain by coincidence would be highly unlikely. However,
more than 50\% of the graphene constrictions display one or several
carbon chains before the two holes merge into one. Therefore, it must
be atomic rearrangements along with the continuous removal of atoms
in the electron beam that leads to carbon chains as the final product,
the thinnest possible carbon bridge. Indeed, several narrow bridges
convert into two parallel chains (Fig. \ref{fig:bigimage}c shows
a rare case of even three) which indicates that this transformation
is energetically favoured rather than caused by only the removal of
atoms.

There is no doubt that these chains must be made from carbon, since
we can follow, atom by atom, the transition from graphene to these
chains (Fig. \ref{fig:bigimage}, and supplementary video). Contrast
and observed width in the experiment is in agreement with the calculation
for single-atomic chains (Figs. \ref{fig:bigimage}f, g). Further,
the amount of carbon material that is present just before the transitions
can not account for more than a single-atomic chain in each of the
dark lines. Several of these chains can occur in parallel (Fig. \ref{fig:bigimage}c,
f) and even convert back to a (2D) planar covalent network (Fig. \ref{fig:bigimage}c
to d). The chain structures may be of the double-bonded (cumulene-type)
or alternating single-triple bonded (polyyne-type) structures \cite{CChainCumuleneVsPolyyneJOSB87,CChainNDRtheoryNL08,CChainCumuleneSci08},
or possibly the linear alkane or polyacetylene type chains. The first
two structures, cumulene (...$=C=C=$...) or poly-yne type (...$-C\equiv C-C\equiv C-$...)
are linear chains and in good agreement with the observation; calculations
show that the number of carbon atoms (even or odd) determines which
case is present \cite{CChainCumuleneVsPolyyneJOSB87,CChainNDRtheoryNL08}.
In our experiment, the signal to noise ratio is not sufficient to
distinguish the two types. The last two structures, an alkane chain
or polyacetylene, are angled chains with bonds at 120° and 109.5°,
respectively. No indication of this zig-zag structure is seen in our
chains; however, its visibility depends on whether the correct projection
occurs in the experiment. If we assume that random orientations occur
in the experiment, then a sufficent number of chains was observed
to rule out these angled types of chains. 

Similar carbon chain configurations have been observed previously
in carbon nanotube samples \cite{CChainAtCntNL03,CChainZettlNL06}
but were often not stable enough for recording a TEM image \cite{CChainZettlNL06,AjayanSurfReconstrPRL98}.
In our case, more than 50\% of the graphene constrictions convert
into a carbon chain at the end of the thinning process. The carbon
chains are frequently stable for one minute, and sometimes up to two
minutes, in our rather intense electron beam. One minute corresponds
to a dose of $\approx2\cdot10^{9}\frac{e^{-}}{\textrm{nm}^{2}}$.
This result is particularly surprising in the light of commonly assumed
radiation dose limits for organic molecules, which are on the order
of $1\cdot10^{4}\frac{e^{-}}{\textrm{nm}^{2}}$ \cite{BioRaddamMalacUltr07},
and thus five orders of magnitude lower than our doses. It has been
speculated that the high conductivity of the graphene sheet reduces
the effect of ionization damage \cite{GraFillChargeReplNJP08}. Of
course, a carbon chain is not a complex organic structure, but it
does represent an organic substance and constitutes a building block
of many more complex molecules. These results indicate that radiation
damage mechanisms are not well understood; experiments at varying
electron energies should help to gain further insight.

\begin{figure}
\includegraphics[width=0.75\linewidth]{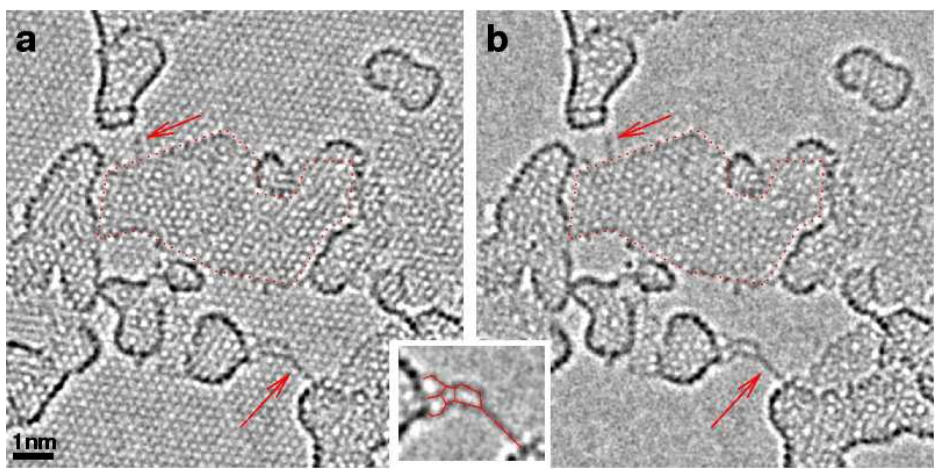}

\caption{Carbon chains suspended between adsorbates on top of a single-layer
graphene membrane. (a) HRTEM image, (b) Fourier filtered image with
the graphene lattice removed. Arrows indicate single-atomic chains
of carbon atoms. One of them branches into a planar covalent carbon
network at one end, as indicated on the inset. The adsorbed {}``contamination''
here has a thickness of only one or two mono-layers: The area indicated
by the red dotted line contains a single-layer network of carbon atoms
on top of the graphene sheet, dominated by 5, 6, and 7-membered carbon
rings, while other areas are at most two layers of contamination.
The chain formation from a constriction in an amorphous adsorbates
is also shown in the supplementary videos S4, S5. \label{fig:cchains}}
\end{figure}

\begin{figure}
\includegraphics[width=0.75\linewidth]{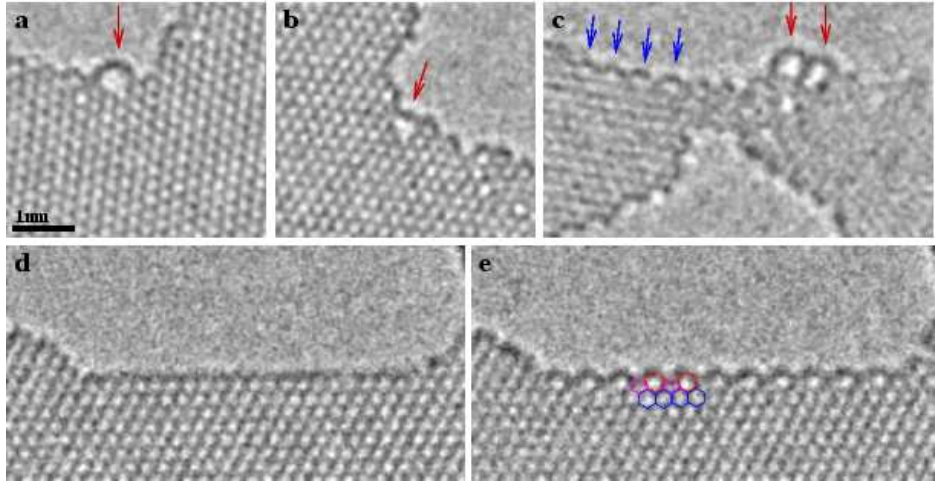}

\caption{Carbon chains observed at the edges of graphene sheets (red arrows).
(a) Individual carbon chain loop at an edge. (b) Carbon chain bridging
a gap at the edge. (c) A double loop configuration. The structure
in (c) is partly vibrating (due to large holes in the membrane) and
therefore the lattice is not well resolved. Also visible in (c) are
a few smaller loops (blue arrows) that indicate a new edge reconstruction.
(d,e) Transition from a zigzag edge to the reconstructed edge, terminated
by pentagons and heptagons. A structural model is overlaid in part
(blue: hexagons, pink: pentagons, red: heptagons.). \label{fig:edgechains}}
\end{figure}

In addition to these free-hanging carbon chains, we observe carbon
chains that are supported by a graphene sheet, as observed before
\cite{MeyerAdatomsNat08}. We find that this type of chain frequently
occurs where amorphous adsorbates shrink under electron irradiation,
and forms similar to the aforementioned type, but now from constrictions
in the contamination layer. The supplementary videos S4, S5 show this
process. Fig. \ref{fig:cchains} shows two carbon chains that are
attached to larger adsorbates at their end points. Some of the adsorbates
are single-layer networks of carbon atoms, as indicated in Fig. \ref{fig:cchains}.
Even though we find a slight preference of these chains to align with
the zig-zag direction of the underlying graphene lattice, again no
indication of a zig-zag shape (as expected for an alkane- or polyacetylene-type
chain) is seen in the carbon chain itself. Thus, we confirm their
structure as either cumulene or poly-yne type as described above.
For the case of chain formation from carbon contamination, it is the
low-contrast background of the graphene support that enables the visualization
of this process by TEM. 

As a third configuration, we observe carbon chains that form loops
along graphene edges. Although more rare than the previously described
chains, they can still be reliably found by observing graphene hole
edges for a longer time. Figure \ref{fig:edgechains} shows three
examples of carbon chains that have separated from the edge. Frequently
they merge back into the graphene edge at a later time. These additional
ways of formation confirm that the carbon chains are a preferred configuration
in a continuously diluted set of carbon atoms, not limited to graphene
constrictions.

Finally, we discuss the edge reconstructions that are shown in Figs.
\ref{fig:edgechains}c,e). Fig. \ref{fig:edgechains}e shows an edge
that is terminated by an alternating sequence of pentagons and heptagons
with a period that is twice as large as that of the zig-zag edge.
It constitutes one of the edge reconstructions predicted in Ref. \cite{GraEdgeReconstPRL08}.
For this edge, the switching between the two configurations was observed
several times as shown in Figs. \ref{fig:edgechains}d,e and in the
supplementary video S6. During TEM observation, all the edges of graphene
are continously changing, due to the energy input from the electron
beam \cite{GraAtEdge08}. However, as this edge switches from a well
defined zig-zag edge to the reconstructed configuration and back multiple
times in-between two exposures, it can not be a coincidental arrangement
formed by random single-atom knock-on events. More likely, these are
two stable configurations, and the energy input from the beam provides
the activation energy that is required to switch between the two cases.
However, this implies that the energy input from the beam has a non-local
effect (possibly via excitation of a phonon) which changes an entire
ca. 5~nm long edge in one step, rather than just kicking individual
atoms to a different position. 

In conclusion, we have shown the transformation from graphene nano-ribbons
to single carbon chains. Planar, $sp^{2}$-bonded networks that deviate
from the hexagonal structure appear as intermediate reconstructions
as the graphene ribbon width shrinks below 1 nanometer. Many electronic
applications of graphene require nanometer-scale graphene ribbons
or graphene constrictions. Thus, on the one hand, reconstructions
have to be considered in ultra-narrow ribbons, while on the other
hand carbon chains might be useful as electronic component and represent
the ultimate constriction in graphene. The chains form efficiently
by self-organization during continuous removal of atoms from a graphene
bridge. Further, they are observed at edges, or form from constrictions
in the carbonaceous contamination. In other words, the {}``precursor''
material for the chains ranges from highly crystalline (graphene)
to amorphous (adsorbate) carbon, which indicates that a rather general
self-organization process is observed here. Finally, the high robustness
under irradiation of these linear carbon chains combined with the
ease of electron beam fabrication at the nm scale can provide a route
to a synthesis of devices and for further studies of these structures.

\bibliographystyle{unsrtnat}
\bibliography{/home/jannik/publ/bibtex/2dstuff,/home/jannik/publ/bibtex/TEM,/home/jannik/publ/bibtex/books,/home/jannik/publ/bibtex/molecules,/home/jannik/publ/bibtex/diverse}

\end{document}